\documentclass[11pt,a4paper]{article}
\usepackage{graphics}
\begin{document}
\hyphenation{brems-strah-lung}
\hyphenation{ki-ne-ma-tics}
\hyphenation{quad-ra-tu-re}
\hyphenation{de-di-ca-ted}
\hyphenation{pre-sent}

\newcommand{\jpsi            }{\mbox{$J/\psi$}}
\newcommand{\xfjpsi            }{\mbox{$x_F(J/\psi)$}}
\newcommand{\chic            }{\mbox{$\chi_c$}}
\newcommand{\rchic            }{\mbox{$R_{\chi_c}$}}
\newcommand{\ccbar           }{\mbox{$c\bar{c}$}}
\newcommand{\qqbar           }{\mbox{$q\bar{q}$}}
\newcommand{\deltaM          }{\mbox{$\Delta M$}}
\newcommand{\mev             }{\mbox{\rm{MeV/$c^2$}}}
\newcommand{\Gev             }{\mbox{\rm{GeV/$c^2$}}}
\newcommand{\ECAL            }{\mbox{ {\sl Ecal}}}
\newcommand{\herab            }{\mbox{HERA-$B$}}

%\begin{titlepage}
\begin{flushright}
DESY-02-187 \\
hep-ex/0211033 \\
20-11-2002
\end{flushright}

\vspace{1.5cm}
\begin{center}
\begin{LARGE}
{\bf \jpsi{} Production  via \chic{} Decays in 920~GeV $pA$ Interactions}
\end{LARGE}
\end{center}
\vspace{0.5cm}
%\titlerunning{ \boldmath{\jpsi{}} Production  via \boldmath{\chic{}} Decays in 920~GeV \boldmath{$pA$} Interactions} 
%\subtitle{The \herab{} Collaboration}
\author{}
%\maketitle

\begin{center}
The \herab{} Collaboration
\vspace{0.5cm}

%\authorrunning{I.~Abt {\it et al.}}
%\author{

I.~Abt$^{28}$,
 A.~Abyzov$^{26}$, 
M.~Adams$^{11}$, 
H.~Albrecht$^{13}$, 
V.~Amaral$^{8}$, 
A.~Amorim$^{8}$, 
S.~J.~Aplin$^{13}$, 
A.~Arefiev$^{25}$, 
I.~Ari\~no$^{2}$, 
M.~Atiya$^{36}$
% ...

\vspace{5mm}
% ...
$^{2}${\it Department ECM, Faculty of Physics, University of Barcelona, E-08028 Barcelona, Spain~$^{b}$} \\
% ...
$^{8}${\it LIP Coimbra and Lisboa, P-3004-516 Coimbra,  Portugal~$^{d}$} \\
% ...
$^{11}${\it Institut f\"ur Physik, Universit\"at Dortmund, D-44227 Dortmund, Germany~$^{c}$} \\
% ...
$^{13}${\it DESY, D-22603 Hamburg, Germany} \\ 
% ...
$^{25}${\it Institute of Theoretical and Experimental Physics, 117259 Moscow, Russia~$^{i}$} \\
$^{26}${\it Joint Institute for Nuclear Research Dubna, 141980 Dubna, Moscow region, Russia} \\
% ...
$^{28}${\it Max-Planck-Institut f\"ur Physik, Werner-Heisenberg-Institut, 
D-80805 M\"unchen, Germany~$^{c}$} \\
% ...
$^{36}${\it Brookhaven National Laboratory, Upton, NY 11973, USA} \\
% ...
$^{b}${\it supported by the CICYT contract AEN99-0483} \\   % Spain 
$^{c}${\it supported by the Bundesministerium f\"ur Bildung und Forschung, FRG, under contract numbers 05-7BU35I, 05-7DO55P, 05 HB1HRA, 05 HB1KHA, 05 HB1PEA, 05 HB1PSA, 05 HB1VHA, 05 HB9HRA, 05 7HD15I, 05 7HH25I,  05 7MP25I, 05 7SI75I } \\
$^{d}${\it supported by the Portuguese Funda\c{c}\~ao para a Ci\^encia e Tecnologia} \\
% ...
$^{i}${\it supported by the Russion Fundamental Research Foundation under grant RFFI-00-15-96584 and the BMBF via the Max Planck Research Award} \\ %ITEP
% ...

%\end{flushleft}

%}
%\input inst.tex
%\date{Received: Revised version: \today }
%\date{\ \ \  \\ \\ \\ \\  \ }
% 
\end{center}
%\end{titlepage}

\vskip 1.0cm 

\begin{abstract} Using data collected by the \herab{} experiment, we have
  measured the fraction of \jpsi{}'s produced via radiative \chic{}
  decays in interactions of $920$~GeV protons with carbon and titanium
  targets.   We obtained
  $ R_{\chi_c} =\ 0.32 \pm 0.06_{stat} \pm 0.04_{sys}$ for the
  fraction of \jpsi{} from \chic{} decays averaged over proton-carbon
  and proton-titanium collisions.  
  This result is in agreement with previous measurements and is
  compared with theoretical predictions.

PACS: 
\hspace{0.5cm} [13.85.Ni], [13.85.Qk], [24.85.+p], [24.85.Eqp]
\end{abstract}
%\PACS{ 
%{13.85.Ni}{~Inclusive production with identified hadrons} \and
%{13.85.Qk}{~Inclusive production with identified leptons, photons, or other
%  non-hadronic particles } \and
%{24.85.+p}{~Quarks, gluons, and QCD in nuclei and nuclear processes} \and
%{25.40.Ep}{~Inelastic proton scattering}
%} % end of PACS codes

%\begin{document}
%\hugehead
%\markboth
%\maketitle
%\twocolumn
%
%
%%%%%%%%%%%%%%%%%%%%%%%%%%%%%%%%%%%%%%%%%%%5
\newpage

%\offprints{}
\section{Introduction}
\label{sec:intro}
%While the production of heavy quark pairs is well described via
%perturbative QCD, 
The mechanism by which quarkonium states are produced in hadronic
collisions
%quarks bind to form quarkonium 
is not understood and is a subject of current interest.  At present,
several  models exist.  The Colour Singlet Model (CSM)
\cite{CSM} requires that the \qqbar{} pair be produced in a colour
singlet state with the quantum numbers of the final meson. The
Non-Relativistic QCD factorisation approach (NRQCD)
\cite{NRQCD,NRQCD1} assumes that a colour singlet or colour octet
quark pair evolves towards the final bound state via exchange of soft
gluons. The nonperturbative part of the process is described by long
distance matrix elements which are extracted from data.  Finally, the
Colour Evaporation Model (CEM) \cite{CEM,CEM2} assumes the exchange of many
soft gluons during the formation process such that the final meson
carries no information about the production process of the \qqbar{}
pair.  
%CEM thus implies that quarkonium formation is independent of
%c.m.s. energy ($\sqrt{s}$) and the projectile/target parton density
%functions.

Charmonium production is an attractive test case as the quarks are
heavy enough for perturbative calculations of the $q\bar{q}$
production process, yet the cross sections are large enough to be
measured with good statistics.  The dependence of the ratio of
production cross sections for different states, e.g. the ratio of
\chic{}\footnote{In the following, the notation ``\chic{}'' indicates
  the sum of the three states 
$\chi_{c0}$, 
$\chi_{c1}$ and
  $\chi_{c2}$.} and \jpsi{} production cross sections
$\sigma(\chi_{c}) / \sigma(J/\psi)$, on $\sqrt{s}$ or the projectile
allows one to distinguish among different
 models.  From the experimental point of
view, the specific decay $\chi_{c} \to J/\psi\,\gamma$ is
advantageous since  
the decay signature
$J/\psi\to\ell^+\ell^-$ ($\ell=\mu,e$)
can be used as trigger requirement.
Furthermore, several
systematic errors cancel in the ratio, and the only significant
difference in the detection of the \chic{} and the \jpsi{}
%two states
is the photon reconstruction. 
Due to the small branching ratio 
of $\chi_{c0} \to J/\psi\,\gamma$,
$(6.6\pm 1.8) \cdot 10^{-3}$ \cite{PDG},
the $\chi_{c0}$ contribution to the reconstructed \chic{} signal can be 
neglected.
%state does not play a significant 
%role due to its small branching ratio $(6.6\pm 1.8) \cdot 10^{-3}$
% \cite{PDG}.
The $\chi_{c1}$ and $\chi_{c2}$ states, with radiative branching 
ratios of $0.273\pm 0.016$ and $0.135\pm 0.011$ \cite{PDG}, respectively, 
are separated by 46 MeV/$c^2$.
In most 
%(but not all) 
experiments the energy resolution is 
insufficient to resolve these two states, so that one 
usually quotes the ratio
\begin{small}
\begin{equation}
  \label{eq:Rchic}
  R_{\chi_c} =
\frac{ \sum\limits_{i=1}^2 \sigma(\chi_{ci})Br(\chi_{ci} \to 
J/\psi\,\gamma) 
%  + \sigma(\chi_{c2})Br(\chi_{c2} \to J/\psi\,\gamma)
}{\sigma(J/\psi)}.
\end{equation}
\end{small}
Here, $\sigma(J/\psi)$ is the sum of production cross sections
for direct \jpsi{}'s and \jpsi{}'s produced in decays of  
\chic{} and $\psi'$. In the same way, $\sigma(\chi_{ci})$ includes
direct \chic{} production and the feed-down from the $\psi'$.
Contributions from
$\eta_c'$, $h_c$ and heavier charmonia are neglected.  

While this ``inclusive'' \rchic{} ratio is usually quoted in the literature,
one can define 
the  ratio for direct \chic{} production over direct 
\jpsi{} production 
\begin{equation}
\label{eq:rchicdir1}
R_{\chi_c}^{dir} =
\frac{\sum\limits_{i=1}^2\sigma(\chi_{ci})_{dir}Br(\chi_{ci}\to 
J/\psi\gamma)}{\sigma(J/\psi)_{dir}}.
\end{equation}
$R_{\chi_c}^{dir}$ can be derived from \rchic{} and the known ratio of
$\psi'$ to \jpsi{} production cross sections \cite{psi} and known
branching ratios \cite{PDG}.

The experimental situation is unclear, and the uncertainties are
large particularly for proton induced reactions  where 
the few existing measurements of \rchic{} \cite{pA}  differ strongly.  
Measurements made with pion beams \cite{piA} have higher
precision and may indicate that \rchic{} increases with $\sqrt{s}$.
For photon and electron-induced reactions, only upper limits for
\rchic{} have been reported \cite{photoproduction}.

We report here a new determination of \rchic{} in interactions of
920 GeV protons with carbon and titanium nuclei.  
The \chic{} is observed in the decay $\chi_c \to J/\psi \gamma \to
\ell^+\ell^-\gamma$ ($\ell=\mu,e$)
using the value
\deltaM{}, which is the difference between the invariant mass of the
($\ell^+\ell^-\gamma$) system and the invariant mass of the lepton pair 
$\ell^+\ell^-$~:
\begin{equation}
  \label{eq:deltaM}
\Delta M = M(\ell^+\ell^-\gamma) - M(\ell^+\ell^-) .
\end{equation}
Here, the uncertainty in the determination of the \jpsi{} mass
essentially cancels.  An excess of events with respect to the
combinatorial background determines the number of \chic{} candidates
$N_{\chi_c}$, from which the ``inclusive'' value, \rchic{}, can be calculated
as follows:
%The value for \rchic{} is calculated via
\begin{equation}
  \label{eq:Rchicmeas}
R_{\chi_c} = \frac{N_{\chi_c}}{N_{J/\psi}\cdot \varepsilon_{\gamma}}  \cdot
\rho_{\varepsilon}  , 
\end{equation}
where $N_{J/\psi}$ is the total number of reconstructed
\mbox{$J/\psi\to\ell^+\ell^-$} decays used for the \chic{}
search. 
The factor $\varepsilon_{\gamma}$ is the
photon detection efficiency.
%The photon detection efficiency is explicitly taken into
%account by the factor $\varepsilon_{\gamma}$.  
The value
$\rho_{\varepsilon}$ represents the ratio of trigger and
reconstruction efficiencies for \jpsi{}'s from \chic{} decays and for
all \jpsi{}'s:
\begin{equation}
  \label{eq:reff}
 \rho_{\varepsilon} = \frac{\varepsilon_{trig}^{J/\psi}}{\varepsilon_{trig}^{\chi_c}}
   \frac{\varepsilon_{reco}^{J/\psi}}
 {\varepsilon_{reco}^{\chi_c}}  .
\end{equation}
Since the kinematics, triggering and reconstruction of 
 direct \jpsi{}'s and \jpsi{}'s from $\chi_{c}$ decays are very
similar, $\rho_{\varepsilon}$ is  close to unity.

\section{Detector, Trigger and Data Sample}

\herab{} is a fixed target experiment operating at the HERA
storage ring at DESY. Charmonium and other heavy flavour states are
produced in inelastic collisions by inserting wire targets into the
halo of the 920~GeV proton beam circulating in HERA. The $pN$
($N=p,n$)
%center-of-mass 
c.m.s. energy is $\sqrt{s} = 41.6$~GeV.  

The detector is a magnetic spectrometer emphasising vertexing,
tracking and particle identification, with a dedicated
\jpsi{}-trigger.
%Dedicated hardware processors allow
%track reconstruction and \jpsi{} selection at the First Level Trigger.
%The experiment  features are a dedicated \jpsi{}-trigger, a high-precision
%tracking system, and several particle identification systems.  
The components of the \herab{} detector used for this analysis include
a silicon strip vertex detector (VDS),
%{\it thsb eliminates ITR, Rich}
%micro-strip gas chambers, 
honeycomb drift chambers (OTR), a large acceptance 2.13 T$\cdot$m magnet,
%a ring imaging \v{C}erenkov counter, 
a finely segmented ``shashlik'' electromagnetic calorimeter (ECAL),
and a muon system (MUON) consisting of wire chambers interleaved with
iron shielding which detects muons with momenta larger than %$\approx$
5 GeV/$c$.  The ECAL is divided into three radial parts with
decreasing granularities, of which two, the ``inner'' and ``middle''
sections, are used for the measurement described here.
% The granularity corresponds
%to 1.6 and 1.34 times the Moli\`ere radius for the inner and the
%middle section, respectively.
The performance of the calorimeter is described in Ref.~\cite{ECAL}.
%{\it The measured energy resolution is $\delta E/E = 23\%/\sqrt{E}
%  \oplus 1.7\%$ for the inner section. For the middle section, the
%  stochastic term of the $\delta E/E $ is $ 13\%/\sqrt{E}$, while the
%  constant term is not measured  \cite{ECAL}.}   
The \herab{} detector allows an efficient reconstruction of particles
with  momenta larger than 1 GeV/$c$, including $\gamma$'s and $\pi^0$'s,
within the acceptance.  A detailed description of the apparatus is
given in Ref.~\cite{herab}.

The \herab{ }target station houses 8  target wires  
which can be moved independently into the beam halo.
Their positions are steered such that the proton interaction rates 
are equalised for the targets in use.
The data presented here were obtained using a %vertical 
carbon wire and a 
%horizontal 
titanium wire separated by 3.3\,cm along the beam direction.  The
resolution of the reconstructed dilepton vertices of 0.6\,mm along the
beam direction \cite{VDS} allows a clear association of the
interaction to a specific target wire. 
%at a proton bunch crossing rate of 10\,MHz. 
%At this interaction rate, the mean
%number of interactions per event was 0.5.
%distributed accordingly to Poisson statistics.
The analysis presented is based on data collected during a short
commissioning run in summer 2000. About half of the data was taken
with a single carbon wire; the second half was taken with carbon and
titanium wires together. The proton-nucleus interaction
rate was approximately 5\,MHz.
% The assignment of an event to a target wire is
%based on the reconstructed dilepton vertex position.

The trigger selects  $\mu^+\mu^-$ and $e^+e^-$  pairs, the latter with
an invariant mass larger than 2 GeV/$c^2$. For a muon candidate the
trigger requires at least 3 MUON hits in coincidence with 9 OTR hits
consistent with a particle track with a transverse momentum between
0.7~GeV/$c$ and 2.5~GeV/$c$.
%within a region of interest
%which is defined by requiring a transverse momentum of muon
%candidates  greater than 0.7~GeV/$c$ and less than 2.5~GeV/$c$. 
The electron trigger requires that the transverse energy deposited in
the calorimeter\footnote{The 
transverse energy is defined as cluster energy
  multiplied by the transverse distance to the beam axis and divided
  by the cluster-target distance.} 
by the electron candidates be greater than 1.0~GeV and
that at least 9 OTR hits confirm the track hypothesis.  Both muon and
electron candidates have to be confirmed by a track segment in the
vertex detector with at least 6 hits.  For the data described here,
the trigger acceptance for \jpsi{}'s was limited to the $x_F$ range
$-0.25<x_F< 0.15$, $x_F$ being the Feynman's $x$ variable. For more
details concerning the trigger and the data sample of the year 2000 run,
see Ref.~\cite{bbar}.
%For the data described here, the \jpsi{} acceptance covers the kinematic range
%$-0.25<x_F< 0.15$, where $x_F$ is the Feynman-x variable.  For more details
%concerning the trigger, see Ref.~\cite{bbar}.

The data is divided into four separate samples: $\mu^+\mu^-$ or $e^+e^-$ 
final states, each originating from either carbon or
titanium target wires ($\mu$-C, $e$-C, $\mu$-Ti and $e$-Ti).

\section{Monte Carlo Simulation}
\label{sec:mc}

At present, NRQCD is the favoured approach to describe charmonium
formation. It is therefore used to generate our signal Monte Carlo
sample.  To estimate the model dependence systematics we use the CSM.
Since the CEM does not make any conclusive predictions for the
differential charmonium production cross sections, we have not used it
in the simulations.

The Monte Carlo simulation  (MC) of events is done in three steps.
First, a \jpsi{} or \chic{} is generated using PYTHIA 5.7 \cite{PYTHIA}.
In the simulations, the CTEQ2L parton
density function \cite{PYTHIA} and the $c$ quark mass $m_c = 1.48$
GeV/$c^2$ are used.
The sum of the transverse momenta of the
reaction products
%the interaction 
must exceed 0.5 GeV/$c$. Any polarisation
is neglected.
For NRQCD, the differential cross sections and long distance matrix
elements are taken from Ref.~\cite{NRQCD1}. For CSM, the
differential cross sections are taken from Ref~\cite{CSM}.
%The transverse momentum and the Feynman's $x$ spectra of \jpsi{}'s and
%\chic{}'s, {\bf as well as their relative ratio} are weighted according to
%NRQCD \cite{NRQCD1} or CSM \cite{CSM}
%models
%neglecting any polarisation.  In the simulations, the CTEQ2L parton
%density function \cite{PYTHIA} and the $c$ quark mass $m_c = 1.48$
%GeV/$c^2$ are used in combination with NRQCD long distance matrix
%elements from Ref.~\cite{NRQCD1}. 
%The sum of the transverse momenta of
%the interaction must exceed 0.5 GeV/$c$.  
During the second step, the
energy remaining after the charmonium generation is used to simulate
the rest of the $pA$ interaction using FRITIOF 7.02 \cite{FRITIOF}.
Finally, the \jpsi{} event is combined with $n$ other inelastic
interactions to simulate several interactions per event, as observed
in the data.  The number $n$ is distributed according to Poisson
statistics with a mean value of~0.5 determined from the mean
experimental interaction rate.
%of the reaction is set to 

The detector response is simulated using GEANT 3.21 \cite{GEANT} and
includes the measured hit resolution, the mapping of inefficient
channels, and electronic noise.  The simulated events are processed by
the same trigger and reconstruction codes as the data.  The simulation
has been checked to ensure that it accurately describes the detector,
both in terms of geometric acceptance and material composition (see
sect. \ref{sec:syst}).  From the MC we expect a mass resolution for
the \chic{} signal of 45 MeV/$c^2$, which is insufficient to separate the
$\chi_{c1}$ and $\chi_{c2}$ states.

MC studies show that the trigger and reconstruction
efficiencies for \mbox{$J/\psi\to\ell^+\ell^-$} are indeed
similar for both direct \jpsi{}'s and for those originating from
\mbox{$\chi_c\to J/\psi\,\gamma$} decays. We obtain
$\rho_{\varepsilon} = 0.95 \pm 0.02$ for the NRQCD
%model 
and $\rho_{\varepsilon} = 0.97 \pm 0.01$ for the CSM.
% model.  
For the measurement we use the NRQCD value $\rho_{\varepsilon} = 0.95$
and consider the difference between the two values as a measure of the
systematic uncertainty of $\rho_{\varepsilon}$ (see
sect.~\ref{sec:syst}).

The Monte Carlo sample used in the analysis is about six times larger
than the data sample.

\section{Data Analysis}

\subsection{Method and Selection Criteria}
\label{sec:gen}

%The analysis consists in the reconstruction of \jpsi{} events, after
%which one looks for photon candidates in the ECAL.  One then
%constructs the \chic{} candidate invariant masses by combining the \jpsi{} and
%all  photon candidates in the event.  
The  analysis consists of the reconstruction of the \jpsi{}
events,  the search for the photon candidates in the ECAL, and  the
determination of the invariant mass of the \jpsi{} and photon
candidates within the event.  The selection criteria for the $\chi_c$
are tuned to maximise the quantity $N_{J/\psi}\cdot
\varepsilon_{\gamma}/\sqrt{N}$. Here, $N_{J/\psi}$ is the number of
$J/\psi$ candidates above background found in the data, and
$\varepsilon_\gamma$ is the photon reconstruction efficiency
determined from MC simulations. $N$ is the number of all events,
including background, found in the measured \deltaM{}
distribution within a window of two standard deviations 
(determined from MC) around the expected
\chic{} position. 
%{\it During the optimisation
%the $N_{J/\psi}$ and $\varepsilon_{\gamma}$ are changed from maximum
%possible value to two thirds of it. }
The procedure is applied for all cuts described below.

%%%%%%%%%%%%%%%%%%%%%%%%%%%%%%%%%%%%%%%%%%%%%%%%%%%%%%%%%%%%%%
\subsection{\boldmath{\jpsi{}} Selection}
\label{sec:jpsi}

In the offline analysis, a track is selected as a muon candidate if
its transverse momentum is greater than 0.7~GeV/$c$ and the muon
likelihood, derived from the ratio
% and the positions 
of the expected
and found MUON hits, is greater than 0.001. The latter removes hit
combinatorics which satisfy the trigger while keeping nearly all good
muons.

A track is identified as an electron candidate if {\it (a)\/} 
the transverse energy ($E_T$) is greater than 1~GeV;
and {\it (b)\/}  it has
$|E/p-1|<0.3$, where $E$ is the energy deposited in the calorimeter
and $p$ is the track momentum.
The cut on $E/p$ corresponds to about 3.3 standard deviations of the
electron $E/p$ distribution.  To further reject the background from
hadrons, a search for associated bremsstrahlung photons emitted in the region
upstream of, or inside, the magnet is performed for each electron
candidate.
%the electron candidates are required to have emitted a bremsstrahlung
%photon in the region upstream of, or inside, the magnet. 
Thus, an isolated electromagnetic cluster is required in the area
where the bremsstrahlung would hit the ECAL.  The energy of the
bremsstrahlung cluster is added to the energy of the electron
candidate.  The requirement of an associated bremsstrahlung photon
candidate for each of the two electron candidates of the
\mbox{$J/\psi\to e^+e^-$} decay has an efficiency $\varepsilon_{brem}$
of about 20\% (about 45\% per electron) and suppresses the background
by a factor of 45.  These values are obtained by comparing the
\jpsi{} and background rates under this requirement with those for the
case that at least one of the two electrons is associated with a
bremsstrahlung cluster, and they are also confirmed by MC studies 
(see alse sect. 4.6).

The assignment of the \jpsi{} candidates to a target wire is based on
the position of the reconstructed dilepton vertex.  The
$\chi^2$ probability of this vertex is required to be larger than
0.005 to eliminate spurious events.

\begin{figure*} %[h]
\begin{center}
\resizebox{0.6\textwidth}{!}{%
  \includegraphics{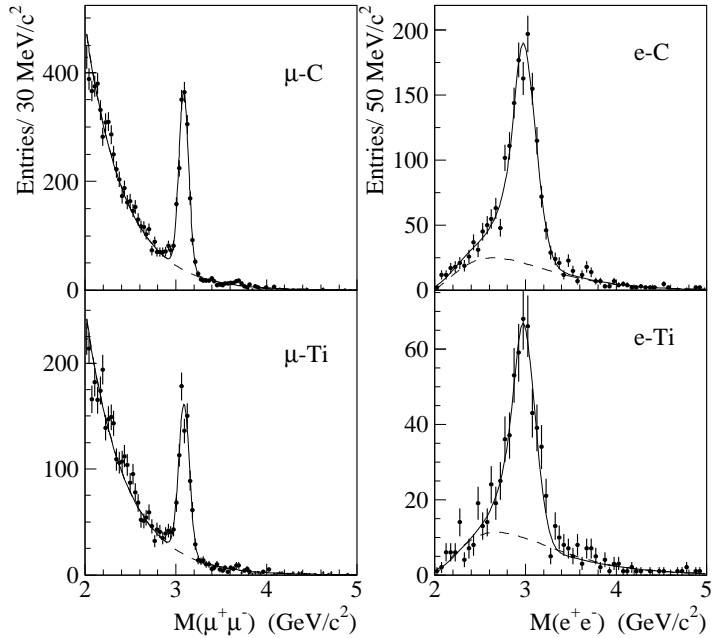}
}
\caption{ Dilepton invariant mass spectrum for each 
  of the four samples ($\mu$-C, $\mu$-Ti, $e$-C, $e$-Ti).  The dashed
  lines show the estimated background under the \jpsi{} signal. See
  text for the details on the fits (solid lines). The selection criteria
  are described in sects.~\ref{sec:jpsi} and \ref{sec:multi}.  }
\label{fig:msJpsi} 
\end{center}
\end{figure*}

The invariant mass is calculated for each opposite-charge lepton pair.
The resulting mass distributions are shown in Fig.~\ref{fig:msJpsi}
for each of the four samples.  The signal observed for $J/\psi \to
\mu^+\mu^-$ events is Gaussian while the $J/\psi \to e^+e^-$ signal
has an asymmetric bremsstrahlung tail. In both cases the background
underneath the signal 
%is mainly combinatorial. 
comes mainly from misidentified hadrons and conversions. 
The background shape
is either described by an exponential distribution ($\mu^+\mu^-$) or
by an exponential multiplied by a second order polynomial distribution
($ e^+e^-$).  The shape is confirmed by the invariant mass
distribution of the same sign candidates in the muon case, and by
fitting the distribution that results from using all trigger
candidates (mostly hadrons) in the electron case.  Only $J/\psi$
candidates within a two standard deviation (2 $\sigma$) window around
the $J/\psi$ mass are considered for the analysis.  In the electron
case, $\sigma$ is taken from the high mass part of the signal which is
Gaussian.  The numbers of $J/\psi$ candidates obtained from the fit
and corrected for the mass window are shown in Table~\ref{tab:results}
for the four samples.

\subsection{Particle Multiplicities}
\label{sec:multi}

%%%%%%%%%%%%%%%%%%%%%%%%%%%%%%%%%%%%%%%%%%%%%%%%%%%%%%%%%%%%%%%%%%%%%%%%%%
Detector occupancies have a considerable impact on the \chic{}
reconstruction: large calorimeter occupancies lead to more
combinatorial background. However, detector occupancies are correlated
with particle multiplicities and thus depend on the underlying physics
of the event.  Since the cross section for \jpsi{} and \chic{}
production are of the same order of magnitude, and the kinematic
dependence of direct \jpsi{} and those from \chic{} decays are
similar, we assume both types of events to have similar particle
multiplicities excluding the decay products of charmonium.  The
charged particle multiplicity is proportional to the number of tracks
reconstructed in the VDS.  We eliminate especially ``busy'' events
which tend to contribute more to the background than to the signal.
Based on the multiplicity distributions we
require that there are not more than 30 (34) VDS tracks in events with
a $J/\psi$ candidate originating from the carbon (titanium) target.
We also require less than 19 clusters in the ECAL.
% with $E>5$~GeV. 
The upper limit on the number of tracks reduces the background under the
\jpsi{} signal while the cut on the number of clusters limits the
combinatorial background under the \chic{}.
%The values of the cuts are
%chosen as stated in sect.~\ref{sec:gen}.
The numbers of \jpsi{}'s passing the multiplicity cuts are given in
Table~\ref{tab:results}.

%%%%%%%%%%%%%%%%%%%%%%%%%%%%%%%%%%%%%%%%%%%%%%%%%%%%%%%%%%%%%%%%%%%%%%%%%%
\subsection{Photon Selection}
Each cluster in the ECAL with $E_T> 0.1$~GeV that is not associated
with the leptons from the \jpsi{}, is considered as a photon candidate.
The area of the ECAL closest to the proton beam pipe $x^2/4 + y^2 <
484$~cm$^2$ (or equivalently~: $\theta_{x}^2/4 + \theta_{y}^2 <
265$~mrad$^2$) is excluded, as the occupancy in this region
is high (up to 30\%).  Hadronic background is reduced by requiring
that the ratio of the central cell energy to the total cluster energy
($E_{centr} /E$) be greater than~0.6.  In order to suppress
background due to soft secondary particles and noise clusters,
%caused by pedestal fluctuations, 
an energy cut $E > 3.0$~GeV is applied.  
%{\it Since the probability of
%  the photon conversion in the OTR in front of the ECAL is large no
%  requirement on the track association with cluster is applied.}
A charged track veto is not applied, due to a 44\% probability of the
photon to convert in the
detector material downstream of the magnet.  The
material of the detector in front of the ECAL causes photon
conversion, and thus losses of photons from \chic{}'s.  We determine
these losses using MC simulations.

Since the relative momenta of the \jpsi{} and the photon 
from the \chic{} decay are correlated,
 a cut in the acceptance of
the \jpsi{} affects the acceptance for the photon as well.  The
electron sample, with a slightly larger acceptance close to the beam
as compared to the muon sample, also contains  more energetic photons
than the muon sample.  The different samples have different
kinematics and acceptances, leading to different photon detection
efficiencies  which are determined for each
sample separately using MC simulations (see
Table~\ref{tab:results}).  The uncertainty in the photon detection
efficiency arises mainly from the finite MC statistics; however, this
uncertainty is insignificant compared to the statistical error on
$N_{\chi_c}$.

%%%%%%%%%%%%%%%%%%%%%%%%%%%%%%%%%%%%%%%%%%%%%%%%%%%%%%%%%%%%%%%%%%%%%%%%%%

\subsection{\boldmath{\chic{}} Reconstruction} 

\begin{figure*}%[h]
\begin{center}
%\strut\psfigure[5cm]{figure2.eps}
\resizebox{0.55\textwidth}{!}{%
  \includegraphics{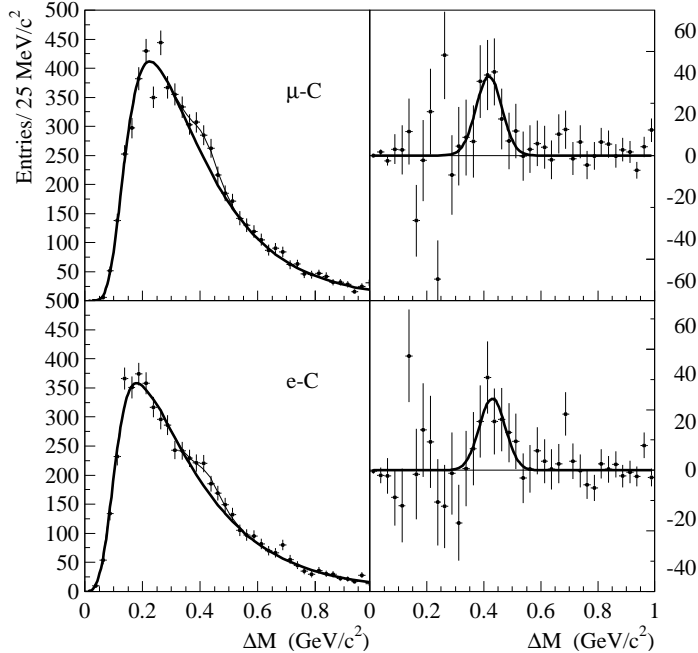}
}
\caption{The $\Delta M = M(\ell^+\ell^-\gamma) - M(\ell^+\ell^-)$ 
  distribution for samples $\mu$-C and $e$-C. In the left-most plots,
  the points represent data and the solid lines represent the
  combinatorial background estimated by event mixing.  The right-most
  plots show the signal after background subtraction. See text and
  Table~\ref{tab:results} for the details on the fit.}
\label{fig:chic} 
\end{center}
\end{figure*}

The \deltaM{} distributions for all combinations of \jpsi{} and photon
candidates for the carbon samples are shown in 
Fig.~\ref{fig:chic}.  The distributions show a signal
%with a resolution of about 45 MeV,
corresponding to the sum of the two charmonium states $\chi_{c1}$ and
$\chi_{c2}$.

Possible sources of background are random combinations of \jpsi{} and
photon candidates, decays of heavier mesons into $J/\psi X$, and the
radiative decay $J/\psi \to e^+e^-\gamma$.  The fraction of \jpsi{}'s
originating from $\psi'$ decays is about 8\% \cite{pA}. The
fraction of photons arising from $\psi'\to J/\psi\,\pi^0 \pi^0$
decays which pass the energy cut $E>3.0$~GeV is negligible, as is the
fraction of pions misidentified as photons from $\psi'\to
J/\psi\,\pi^- \pi^+$ decays. The fraction of \jpsi{}'s resulting from decays
of $B$-mesons, $\Upsilon$, $\chi_b$, $\eta_c'$, and $\chi_{c0}$ is
negligible as well.  The photon from the radiative decay $J/\psi \to
e^+e^-\gamma$, mostly oriented along the direction of one of the
leptons, is indistinguishable from bremsstrahlung and, as such, is taken
into account.  In the muon sample, bremsstrahlung clusters are
neither expected nor  found above background. The fraction of such
radiative decays is considered to be negligible. Thus the background
consists mainly of random combinations of \jpsi{} and photon
candidates.

The shape of the dominantly combinatorial background in the \deltaM{}
distribution is obtained by combining \jpsi{} candidates with photon
candidates from different events with similar multiplicity and
applying the standard selection cuts.  These ``mixed events''
reproduce the shape of the $\Delta M$ distribution everywhere except
in the \chic{} signal region (see solid line in Fig.~\ref{fig:chic},
left panel).  Similar results are obtained when events in the
sidebands of the \jpsi{} mass region are combined with photon
candidates. Since the experimental resolution is of the same order as
the mass difference between $\chi_{c1}$ and $\chi_{c2}$ states and the
statistics is limited, we use a single Gaussian to describe the
signal. In the fit, the position and normalisation of the Gaussian, as
well as the normalisation of the background, are left free.  The width
of the Gaussian is fixed to the value predicted by MC based on the NRQCD
approach (45 MeV/$c^2$), where the production cross section ratio of
$\chi_{c1}$ and $\chi_{c2}$ is approximately 0.65.  The position of
the Gaussian agrees well with the value expected from MC.
%(420 MeV/$c^2$).  
The background normalisation is also treated as a free
parameter when we fit the number of $\chi_{c} \to J/\psi\,\gamma$
decays. The background subtracted distributions are shown in the right
panel of Fig.~\ref{fig:chic}.  The significance of the signals seen in
the $\mu$-C and $e$-C samples is about three standard deviations.  The
obtained number of \chic{} events as well as the number of \jpsi{}
events and the photon detection efficiency are summarised in
Table~\ref{tab:results}.

\begin{figure*}%[h]
\begin{center}
%\strut\psfigure[5cm]{figure3.eps}
\resizebox{0.55\textwidth}{!}{%
  \includegraphics{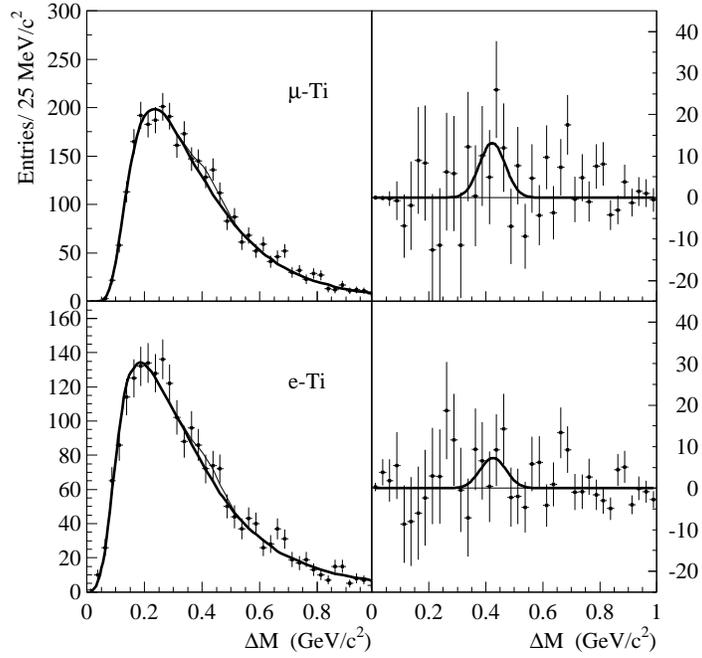}
}
\caption{
  Same as in Fig.~\ref{fig:chic} for the  $\mu$-Ti and $e$-Ti samples.}
\label{fig:chic2} 
\end{center}
\end{figure*}

Taking into account the high background level and the ratio of
$N_{\chi_c}$ to $N_{J/\psi}$ observed in the carbon samples, we do not
expect to see a significant \chic{} signal in the smaller titanium
samples.  The results obtained for the titanium sample with the same
procedure are shown in Fig.~\ref{fig:chic2}. The starting value for
the peak position has been taken from the fit of the carbon data.
Although the signals are marginal, the \rchic{} values obtained from
them are
%nevertheless 
compatible with estimates
from the carbon samples (see Table~\ref{tab:results}).

%The statistics of the Ti-samples
%are much less and do not allow to clearly establish a \chic{} signal
%as expected from the lower number of \jpsi{} events.
%, the statistical significance of
%the excess of events being only two  ($\mu$-Ti) and one ($e$-Ti)
%standard deviations.

\begin{figure*}%[h]
\begin{center}
%\strut\psfigure[5cm]{figure4.eps}
\resizebox{0.55\textwidth}{!}{%
  \includegraphics{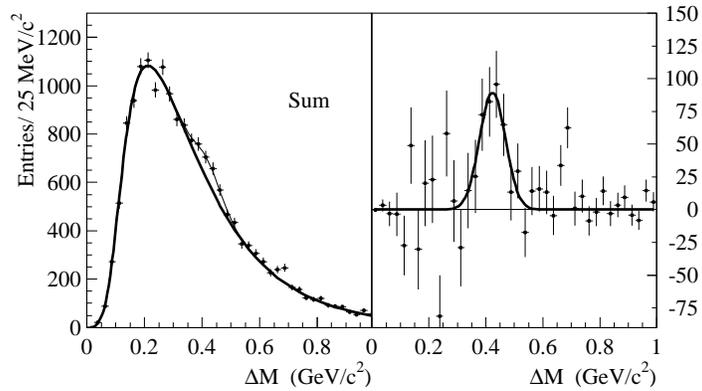}
}
\caption{  Same as in Fig.~\ref{fig:chic}  for all data combined. }
\label{fig:chic3} 
\end{center}
\end{figure*}

As a cross check, all four samples are combined together as shown
in Fig.~\ref{fig:chic3}. The value of $N_{\chi_c} = 380 \pm 74$
obtained from this distribution agrees within the errors with the sum
of the $N_{\chi_c}$ values obtained from the individual samples.

\begin{table*}
  \begin{center} 
\caption{The number of selected \jpsi{} events 
  ($N_{J/\psi}^{selected}$), the number of \jpsi{}'s passing the
  multiplicity cut ($N_{J/\psi}$), the number of \chic{}'s observed
  ($N_{\chi_c}$), $\chi^2$ per degree of freedom for the \deltaM{}
  fit, photon detection
  %probability 
  efficiency ($\varepsilon_{\gamma}$), and the result for
  $R_{\chi_c}$, for each of the four event samples.  The quoted error on
  \rchic{}  is statistical, excluding the systematic uncertainty in
  $\varepsilon_{\gamma}$. }
%and the second one is due to the uncertainty in $\varepsilon_{\gamma}$. }
  \label{tab:results} 
      \vspace{0.3cm}
\begin{tabular}{lcccc}
\hline\noalign{\smallskip} 
                         & $\mu$-C     & $e$-C  &  $\mu$-Ti  &  $e$-Ti \\
\noalign{\smallskip}\hline\noalign{\smallskip}
$N_{J/\psi}^{selected}$  & $1760\pm 48   $&$1380 \pm 69   $&$765\pm 31  $&$512\pm 41$  \\

%$N_{J/\psi}$             & $1516\pm 39 $&$647\pm 25  $&$1180 \pm 83 $&$382\pm 40$  \\
%$N_{\chi_c}$             & $165 \pm 47 $&$ 64\pm 34  $&$121\pm 38 $&$ 31 \pm 27$  \\
%$\chi^2/n.d.f$           & $ 28/35     $&$  27/35 $&$34/35 $&$ 48/35$  \\
%$\varepsilon_{\gamma}$,\%& $26.5\pm 1.1$&$27.1\pm 1.8$&$32.8\pm 1.5$&$ 32.7 \pm 2.6$\\
\noalign{\smallskip}\hline
%$R_{\chi_c}$       & $0.39\pm 0.11   $&$0.35\pm 0.18 $  &$ 0.30\pm 0.10$ & $
%0.23\pm 0.21$\\

$N_{J/\psi}$             & $1510\pm 44      $&$1180 \pm 59           $&$643\pm 29            $&$382\pm 32$  \\
$N_{\chi_c}$             & $159 \pm 47      $&$121\pm 38             $&$ 59\pm 33            $&$ 31 \pm 27$  \\
$\chi^2/n.d.f$           & $ 28/35          $&$34/35                 $&$  27/35              $&$ 48/35$  \\
$\varepsilon_{\gamma}$ (\%)& $27.3\pm 1.1     $&$32.8\pm 1.5           $&$24.4\pm 1.8          $&$ 32.7 \pm 2.6$\\
\noalign{\smallskip}\hline
%$R_{\chi_c}$       & $0.37\pm 0.11 \pm 0.01 $&$0.36\pm 0.20 \pm 0.03$&$ 0.30\pm 0.10\pm 0.01 $&$0.23\pm 0.21\pm 0.02$\\
$R_{\chi_c}$       & $0.37\pm 0.11 $        &$ 0.30\pm 0.09$&$0.36\pm 0.20 $&$0.23\pm 0.21$\\
\noalign{\smallskip}\hline
    \end{tabular}
  \end{center}
\end{table*}

\subsection{Study of the Systematic Uncertainties}
\label{sec:syst}

The systematic uncertainty in the yield of $J/\psi\to
e^+e^-$ candidates due to the background description is  
estimated to be 5\%, whereas the uncertainty
is negligible in the muon case. 

The model dependence of the relative efficiency $\rho_{\varepsilon}$
for all \jpsi{}'s and \jpsi{}'s from \chic{} has been studied. A 2\%
difference of $\rho_{\varepsilon}$ is found for the two models (NRQCD
and CSM) for each of the \jpsi{} leptonic decay modes. For the same models
a difference of the photon detection efficiency $\varepsilon_{\gamma}$
integrated over all \chic{} states is found to be 4\%.  In both cases
the observed difference is treated
%conservatively 
as an estimate of the corresponding systematic uncertainty.  The
overall systematic error accounting for the model dependence of the
selection efficiency is 5\%.

To confirm the MC description of the detector material composition and
acceptance which affects the photon detection
%probability 
efficiency, we compare the bremsstrahlung tag probability
$\sqrt{\varepsilon_{brem}}$ determined from the data and MC.  The
values obtained, $0.44\pm 0.02$ and $0.43\pm 0.01$ for the data and
MC, respectively, are compatible within one standard deviation.  The
systematic uncertainty due to photon losses is conservatively taken to
be~2\%.

The effect of systematic uncertainties of the ECAL calibration is
studied by using MC simulations. The  level of  possible
uncertainties is determined from the data using the $\pi^0$ signal.
The uncertainty of the $\pi^0$ calibration is used then in the MC simulations 
to determine its effect on the detection of the \chic{}.  The
systematic uncertainty on \rchic{} due to this effect is estimated to
be~1\%.

Correlated electronic noise in the calorimeter can shift and widen the
$\Delta M$ distribution of the \chic{} signal. A cluster
reconstruction procedure based on the known correlation between
channels is developed to compensate for this effect.  The numbers of
events observed with and without this algorithm in the data agree with each other
within the statistical errors.  MC studies show that the
corresponding relative systematic uncertainty is 3\%.

The width of the \chic{} signal in the $\Delta M$ distribution depends
on the ratio of $\chi_{c1}$ and $\chi_{c2}$ and on the detector
resolution, mainly that of the ECAL. 
%When we perform the fit for the carbon
%data and allow the width to vary
When the width is left free to vary in the carbon data fit,
the resulting width agrees within
one standard deviation with the nominal one.
%of the two states, on the precision of the ECAL
%calibration, and on the level of electronic noise in the ECAL.  
A systematic error on \rchic{} of 6\% is assigned based on an MC study
of the signal resolution dependence on the $\chi_{c1}$ to $\chi_{c2}$
ratio and the ECAL resolution.
%  {\it comparing the ratio of
%  $\chi_{c1}$ and $\chi_{c2}$ in the NRQCD and the CSM and varing the
%  ECAL resolution in the MC.}

The stability of our results with respect to variations in the
selection criteria is studied separately for the different samples.
The ratio \rchic{} is measured as a function of the cuts on VDS track
multiplicity, ECAL cluster multiplicity, photon energy $E$, and
ratio $E_{centr}/E$.  The variation of the cut on the photon energy
$E$ results in a
%maximum 
variation of \rchic{} of 6\%, which is taken as an estimate for
the systematic uncertainty. The dependence on other cuts is
negligible.

%The only uncertainty comes from the cut  on $E$ and does not exceed 6\%
%The systematic uncertainty due to the applied cuts was
%determined from the resultant variation in \rchic{}.  
%The dependence of \rchic{} on these cuts was found to be~6\%,

The systematic uncertainty on $\varepsilon_{\gamma}$  due to
the finite MC statistics is 3\% as estimated from the weighted 
average of the values in Table~\ref{tab:results}.

The polarisation of the $\chi_c$ affects the reconstruction efficiency
of $\chi_c$, however with the present statistics we are not able to
determine it.  Thus, like the previous experiments, we have assumed no
polarisation and have neglected the uncertainty related to it.

%For the extreme cases for the photon angular distribution in the \chic{}
%rest frame with respect to the \chic{} lab momentum of $\cos^2\theta$ and $\sin^2\theta$,
%\rchic{} takes the values of 0.54 and 0.26, respectively.

Assuming that all individual systematic errors are uncorrelated, an
estimate of the total systematic uncertainty on \rchic{} is
%given by their  quadratic sum  of 
11\% 
%for dimuon sample and 12\% for dielectron samples. 
(see Table~\ref{tab:system}).

\begin{table}[htbp]
  \begin{center}
    \caption{ Contributions to the relative systematic uncertainty.
      For the calculation of the total uncertainty the correlation in
      the systematic errors of the different samples is taken into
      account. }
      \vspace{0.3cm}
    \begin{tabular}{lc}
\hline\noalign{\smallskip} 
  & Uncertainty \\
\noalign{\smallskip}\hline\noalign{\smallskip}
\jpsi{} background shape ($e$-C, $e$-Ti only)    & 5\% \\
Model dependence                                 & 5\% \\
Photon losses                                    & 2\% \\
ECAL calibration                                 & 1\%\\
Correlated noise in ECAL                         & 3\% \\
\deltaM{} resolution                             & 6\% \\
Dependence on cuts                               & 6\% \\
Finite MC statistics ($\varepsilon_{\gamma}$)    & 3\% \\
\noalign{\smallskip}\hline
Total                                            & 11\% \\
\noalign{\smallskip}\hline
\end{tabular}
    \label{tab:system}
  \end{center}
\end{table}

%%%%%%%%%%%%%%%%%%%%%%%%%%%%%%%%%%%%%%%%%%%%%%%%%%%%%%%%%%%%%%%%%%%

\section{Results}

The values of $R_{\chi_c}$ obtained for all four samples are listed in
Table~\ref{tab:results}. The results for the two carbon samples 
agree with each other within the statistical errors.
The results obtained from the titanium data
are consistent with those obtained from the carbon data.
%The weighted average of the carbon results ($\mu$-C and $e$-C) is:
%\begin{equation}
%  \label{eq:averC}
%  \langle R_{\chi_c}^C\rangle\ =\ 0.336 \pm 0.072_{stat} \pm 0.037_{sys} . 
%\end{equation}
%and for the titanium ($\mu$-Ti and $e$-Ti):
%\begin{equation}
%  \label{eq:averTi}
%  \langle R_{\chi_c}^{Ti}\rangle\ =\ 0.297 \pm 0.138_{stat} \pm 0.034_{sys} .
%\end{equation}

%On the basis of models that predict a nuclear dependence 
%of \chic{} and \jpsi{} production cross sections \cite{Vogt2}, 
%one expects, in $pN$ collisions, an increase of $R_{\chi_c}^C$ 
%with respect to $R_{\chi_c}^p$ of up to 5\%,
%and for $R_{\chi_c}^{Ti}$ of up to 8\%. 
%Clearly, the present statistical uncertainties on the
%measured values preclude any quantitative test os  these predictions.
Although nuclear dependence effects might be present in \rchic{}
at the few percent level for the targets used here \cite{Vogt2},
they are beyond the statistical accuracy of the present
 measurement.
%From theory one expects $R_{\chi_c}$ to have a nuclear dependence 
%\cite{Vogt2}. However, in this paper the expected relative difference
%for the two nuclei used is at most a few percent.  Within the
%statistical uncertainties, there is no difference of our results for
%$R_{\chi_c}^C$ and $R_{\chi_c}^{Ti}$.  
%carbon and titanium. 
We therefore average the results for the four samples obtaining:
\begin{equation}
  \label{eq:averTot}
%  \langle R_{\chi_c}\rangle\ =\ 0.321 \pm 0.064_{stat} \pm 0.035_{sys} .
  \langle R_{\chi_c}\rangle\ =\ 0.32 \pm 0.06_{stat} \pm 0.04_{sys} .
\end{equation}
The first uncertainty listed is statistical only, whereas the second
uncertainty is systematic.
%The latter includes also the uncertainty in
%$\varepsilon_{\gamma}$ due to the finite MC statistics.

In order to extract the ratio $R_{\chi_c}^{dir}$ of the ``direct''
\chic{} and \jpsi{} production, we use
\begin{equation}
R_{\chi_c}^{dir} =
\frac{1-R_{\psi}B_1}{1-R_{\chi_c}-R_{\psi}B_1+R_{\psi}B_2} - 1,
%= 0.50 \pm 0.12
\end{equation}
where $R_{\psi} =
\sigma(\psi')/\sigma(J/\psi)
= 0.094\pm 0.035$ is taken from Ref.~\cite{psi} 
%= 0.0165\pm 0.0020$ \cite{psi}, 
and corrected for the branching ratios \cite{PDG}.  $B_1$ is the
branching ratio $ Br(\psi'\to J/\psi X)$ and $B_2$ is the sum of
branching ratios $\sum\limits_{i=1}^2 Br(\psi'\to
\chi_{ci}\gamma)Br(\chi_{ci}\to J/\psi\gamma)$ \cite{PDG}.  We obtain
$R_{\chi_c}^{dir} = 0.50 \pm 0.15_{stat} \pm 0.08_{sys}$.

  \begin{table*}[tp]
    \begin{center}
      \caption{Previous $\pi A$ \cite{piA}, $p A$ \cite{pA}, $p\bar{p}$ 
        \cite{CDF} and \herab{} measurements of the \rchic{} value.
        The value quoted for exp. E771 has been calculated from the
        published cross sections \cite{pA} and branching ratios
        \cite{PDG}.}
      \vspace{0.3cm}
      \label{tab:ratio}
      \begin{tabular}{lccc}
        \hline
Exp. & coll.&$\sqrt{s}$ (GeV) & \rchic{} \\
        \hline
IHEP140 & $\pi^- p$ & 8.5 & $ 0.44\pm 0.16$  \\
WA11  &$\pi^-$Be & 18.7 & $ 0.30\pm 0.05$ \\
E610  &$\pi^-$Be & 18.9 &$0.31 \pm 0.10$ \\
E673  &$\pi^-$H$_2$, $\pi^-$Be & 20.2 &$0.70 \pm 0.28 $ \\
E369  &$\pi^-$Be & 20.6 & $ 0.37\pm 0.09$ \\
E705  &$\pi^-$Li & 23.8 & $ 0.37\pm 0.03$ \\
E705  &$\pi^+$Li & 23.8 & $ 0.40 \pm 0.04$  \\

E672/706  &$\pi^-$Be& 31.1&$0.443 \pm 0.041\pm 0.035$\\
%$ 0.026\pm0.006$  &$0.443 \pm 0.054$&$ 0.57\pm0.19$ \\
        \hline
E610  &$p$Be & 19.4, 21.7 &$ 0.47\pm0.23 $ \\
E705  &$p$Li & 23.8     &$ 0.30\pm 0.04$ \\
E771  &$p$Si& 38.8       &$0.74 \pm 0.17$       \\
R702  &$p p$ & 52, 63     &$ 0.15_{-0.15}^{+0.10} $  \\
ISR  &$p p$ & 62  &$ 0.47\pm 0.08 $  \\
\hline
CDF & $p \bar{p}$ & 1800  & $ 0.297\pm0.017\pm0.057$ \\
\hline
HERA-$B$ & $p$C, $p$Ti & 41.6  & $ 0.32\pm0.06\pm0.04$ \\
%$0.034 \pm 0.01$ &$ 0.297\pm0.059$ &$ 0.96\pm 0.29 $\\
        \hline
      \end{tabular}
    \end{center}
  \end{table*}
  
\begin{figure*}%[3]
\begin{center}
%\strut\psfigure[7cm]{figure5.eps}
\resizebox{0.60\textwidth}{!}{%
  \includegraphics{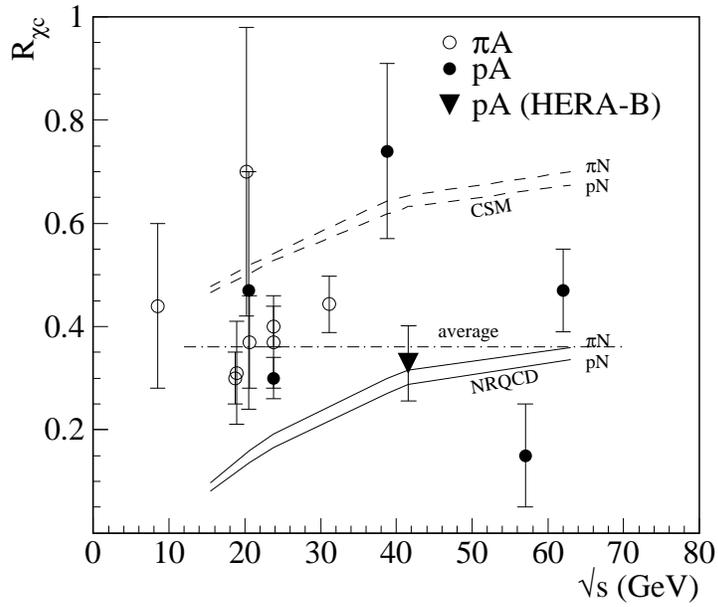}}
\caption{Comparison of our measurement of $R_{\chi_c}$ 
  (closed triangle) with those of other $pp$, $pA$ \cite{pA} (closed
  circles) and $\pi p$, $\pi A$ \cite{piA} (open circles) experiments.
  The CDF result \cite{CDF} is not shown, since its kinematic
  acceptance differs strongly from the other experiments. The value
  quoted for exp. E771 has been calculated from the published cross
  sections \cite{pA} and branching ratios \cite{PDG}. The error bars
  include statistical and systematic uncertainties.  Also shown are
  predictions for $pN$ and $\pi N$ interactions obtained from
  Monte Carlo based on the NRQCD
  \cite{NRQCD1} (solid), CSM \cite{CSM} (dashed) (see
  sect.~\ref{sec:mc}).  The CEM \cite{CEM,CEM2} predicts a constant value.
  The dot-dashed line is the average of all  measurements. }
\label{fig:results}
\end{center} 
\end{figure*}

Our result for $R_{\chi_c}$ (eq. (6))
is compatible with most of the previous data (\cite{pA,piA} 
and \cite{CDF}), as shown in Table~\ref{tab:ratio} and Fig.~\ref{fig:results}.
Due to the relatively large uncertainties, especially for the data on
proton induced reactions, %almost all data are compatible with 
a flat energy dependence, as predicted by CEM \cite{CEM2}, 
cannot be ruled out. 
Similarly, the slope of the energy dependence as
predicted by the Monte Carlo based on NRQCD  (see sect.
\ref{sec:mc})
is also compatible with the data. 
%(except the point at lowest energy).
%, and also our result is compatible with such a behaviour. 
However, the predictions of NRQCD seem to fall below the data,
which might indicate that
%Proton and pion data together indicate that 
the NRQCD long distance matrix
elements for \chic{} production \cite{NRQCD1} used for the calculations
are underestimated.
On the other hand, CSM predicts values for $R_{\chi_c}$  
which are larger than most of the data.
%, but more data are needed to
%discriminate among these models.
More precise measurements, especially of proton induced reactions
 are needed to conclusively discriminate
among these models.

%Due to
%the improved trigger efficiency of \herab{} it is expected that future
%data will significantly improve the knowledge on the value of
%$R_{\chi_c}$.

%%%%%%%%%%%%%%%%%%%%%%%%%%%%%%%%%%%%%%%%%%%%%%%%%%%%%%%%%%%%%%%%%%

\section{Conclusions}

A measurement of the
%The measured 
%ratio of the production cross sections for \chic{} and
ratio of \jpsi{} produced via radiative \chic{} decays 
to all produced  \jpsi{}
allows one to
quantitatively test different models of quarkonium production.  
We present a new result from the \herab{} experiment for the
fraction of \jpsi{}'s originating from radiative decays of $\chi_{c1}$
and $\chi_{c2}$ states produced in $p$C and $p$Ti interactions.
%For the weighted average result we obtain:
%\begin{equation}
%   R_{\chi_c} =\ 0.321 \pm 0.064_{stat} \pm 0.035_{sys} .
%\end{equation}
The fraction of \jpsi{}'s in the range of $-0.25 < x_F < 0.15$ 
originating from radiative \chic{} decays is determined to be 
$R_{\chi_c} =\ 0.32 \pm 0.06_{stat} \pm 0.04_{sys}$,
and consequently, the
ratio of the cross section of directly produced \chic{}'s 
decaying into
\jpsi{} to the cross section of directly produced \jpsi{}'s 
is $R_{\chi_c}^{dir} = 0.50 \pm 0.15_{stat} \pm 0.08_{sys}$
in the above $x_F$ range.
The result has been obtained with C and Ti targets and detecting the
\jpsi{} decay modes into electrons and muons.
Our result for $R_{\chi_c}$ agrees with most previous
proton and pion beam measurements, 
neglecting any possible energy dependence.
It agrees also with the predictions of
the Non-Relativistic QCD factorisation approach (NRQCD), 
whereas it falls significantly % about two standard deviations 
below the predictions of the Color Singlet Model (CSM).
%model.

%%%%%%%%%%%%%%%%%%%%%%%%%%%%%%%%%%%%%%%%%%%%%%%%%%%%%%%%%%%%%%%%%%%%%%%%%%

\section*{Acknowledgements}
\setcounter{footnote}{2}
\renewcommand\thefootnote{\fnsymbol{footnote}}

We express our gratitude to the DESY laboratory and to the DESY
accelerator group for their strong support since the conception of the
\herab{} experiment. The \herab{} experiment would not have been
possible without the enormous effort and commitment of our technical
and administrative staff. It is not possible to list here the many
individuals who have contributed to \herab{}. 
%We are especially
%grateful to the following persons and their groups: G.~Avoni,
%%%%%C.~Baldanza, H.~Bertelsen, J.~Bizzell, A.~Cotta-Ramusino, F.~Czempik,
%I.~D'Antone, J.~Davila, J.~Dicke, A.~Donat, U.~Dretzler,
%A.~Epifantsev, S.~Fricke, W.~Funk, A.~Gutierrez, F.~Hansen, M.~Harris,
%S.~Hennenberger, J.~Hogenbirk, M.~Jablonski, V.~Kiva, M.~Kolander,
%Y.~Kolotaev, L.~Laptin, H.~Leich, H.~L\"udecke, Q.~Li, K.~Y.~Liu,
%P.~Liu, C.~Lu, K.~Ludwig, J.~McGill, E.~Michel, N.~Murthy, E.~Novikov,
%S.~Omeltchuk, D.~Padrazo, H.~B.~Peters, P.~Pietsch, M.~Pohl,
%N.~Ratnikova, A.~Rausch, W.~Reinsch, P.~Rose, I.~Rostovtseva, R.
%Rusnyak, W.~Sands, P.~Solc,
%S.~Starostin$^{\dagger}$\footnote[0]{$^\dagger$ \it deceased },
%K.-H.~Sulanke, V.~Tchoudakov, M.~Tkatch, K.~Wagner, P.~Wegner,
%V.~Zerkin, E.~Zimmer-Nixdorf. We thank the external \herab{} referees
%R. Forty, D.~Froidevaux, \mbox{R.-D.~Heuer}, K.~Jakobs and J.~Jaros
%for many stimulating discussions and suggestions. We are indebted to
%our administrative staff U.~Djuanda and I.~Kerkhoff for their
%continuous assistance.
In the preparation of this paper, we have benefited from many useful
discussions with A.~Leibovich and J.~Lee on the theory of
heavy quarkonium production.

%The authors are grateful to A.Leibovich and J.Lee for the discussions
%and the assistance with the implementation of NRQCD for the MC
%simulations.

\end{document}